\documentclass[aps,prl,twocolumn,showpacs,superscriptaddress,groupedaddress]{revtex4}  

\usepackage{graphicx}  
\usepackage{dcolumn}   
\usepackage{bm}        
\usepackage{amssymb}   
\usepackage{latexsym,bm}
\usepackage[tbtags]{amsmath}

\begin{document}

\widetext


\title{Note on Divergence of the Chapman-Enskog Expansion for Solving Boltzmann Equation}

\author{Nanxian CHEN}
\affiliation{State Key Laboratory of Low-dimensional Quantum Physics and Department of Physics\\
Tsinghua University, Beijing, China}

\email{nanxian@mail.tsinghua.edu.cn}

\author{Bohua SUN}

\affiliation{%
Department of Mechanical Engineering\\
Cape Peninsula University of Technology, Cape Town, South Africa
}%

\email{sunb@cput.ac.za}


\begin{abstract}
\small
Within about a year (1916-1917) Chapman and Enskog independently proposed an important expansion for solving the Boltzmann equation. However, the expansion is divergent or indeterminant in the case of relaxation time $\tau \geq 1$. Even since this divergence problem has puzzled this subject for a century. By using a modified M\"obius series inversion formula, this paper proposes a modified Chapman-Enskog expansion with a variable upper limit of the summation. The new expansion can give not only a convergent summation but also provide the best-so-far explanation on some unbelievable scenarios occurred in previous practice.
\end{abstract}

\maketitle

The Boltzmann equation, developed by Ludwig Boltzmann in 1872 \cite{landau,chapman,cer}, describes the statistical behaviour of a thermodynamic system in a non-equilibrium state. To solve the Boltzmann equation, some attempts have been tried to simplify the collision term in the equation. So far, the most popular scheme is proposed by Bhatnagar, Gross and Krook \cite{bgk,chapman,cer} who made the assumption that the effect of molecular collisions is to force a non-equilibrium distribution function at a point in physical space back to a Maxwellian equilibrium distribution function and at the rate that is proportional to the molecular collision frequency. Therefore, the Boltzmann equation can be modified to the following equation
\begin{equation}\label{e01}
  \frac {\partial f}{\partial t}+{\frac {\mathbf {p} }{m}}\cdot \bm \nabla f+\mathbf {F} \cdot {\frac {\partial f}{\partial \mathbf {p} }}=\frac{g-f}{\tau}.
\end{equation}
Eq.(\ref{e01}) is called the Bhatnagar-Gross-Krook equation (BGK equation in short), in which $f$ is distribution function, $\bm F$ is external force£¬$m$ is mass£¬$\bm v$ is velocity field, $p=m\bm v$ is momentum field, $\tau $ is relaxation time, $g$ is the local Maxwellian distribution function given the gas temperature at this point in space. The relaxation time is defined as $\tau \sim \frac{\lambda}{L}K_n$, where $\lambda$ is the molecular mean free path£¬$L$ is the characteristic length£¬$K_n=\lambda/L$ is the Knudsen number \cite{landau,chapman,cer}.

If not considering the force $\bm F$, Eq. (\ref{e01}) is reduced to
\begin{equation}\label{e02}
  g(t)=f(t)+\tau \frac{Df}{Dt}.
\end{equation}
where the operator $\frac{D}{Dt}=\frac {\partial }{\partial t}+\bm v \cdot \bm \nabla $.

Assume local the Maxwellian distribution function $g(t)$ is given, it follows from Eq. (\ref{e02}) that the distribution function $f(t)$ can be constructed in terms of  $g(t)$ using formal expansion
\begin{equation}\label{e03}
\begin{split}
f(t)&=g(t)-\tau\frac{Df}{Dt}\\
&=g-\tau\frac{D}{Dt}(g-\tau\frac{Df}{Dt})\\
&=g-\tau\frac{Dg}{Dt}+\tau^2\frac{D^2g}{Dt^2}-\cdots.
\end{split}
\end{equation}
or equivalently
\begin{equation}\label{e04}
\begin{split}
f(t)&=\sum \limits_{n=0}^\infty(-1)^n\tau^n\frac{D^n}{Dt^n}g(t).
\end{split}
\end{equation}
This expansion is called the Chapman-Enskog expansion proposed by Enskog and Chapman independently \cite{chapman,cer}.

Unfortunately, the summation in Eq. (\ref{e04}) is divergent or indeterminate under $\tau \geq 1$, hence, the Chapman-Enskog expansion is just a formal-solution which does not guaratee convergence. To ensure convergence in different parameter regimes, many efforts have been made but significant results for a unified approach are far from satisfactory \cite{bgk,chapman,cer,mc,santo}.

In mathematics, a divergent series is an infinite series that is not convergent, meaning that the infinite sequence of the partial sums of the series does not have a finite limit. If a series converges, the individual terms of the series must approach zero. Thus any series in which the individual terms do not approach zero diverges. With this understanding, if assuming the derivative $\lim \limits_{n \rightarrow\infty}\frac{D^n}{Dt^n}g(t)$ is bounded, then $\sum \limits_{n=0}^\infty(-1)^n\tau^n\frac{D^n}{Dt^n}g(t)$ will be divergent or indeterminate under $\tau \geq 1$. To handle the divergence in the case of $\tau \geq 1$, it is understood that the summation should not be made up to infinity instead of a finite number of terms, namely, the series summation must be truncated at a finite number of terms. For a divergent series, the question is how to determine the finite number of truncated terms. From convergence study of an infinite series, we know that the convergence is not only depending on the independent expansion variable $\tau$ but also its maximum $\tau_{max}$, and related to their ratio $\tau_{max}/\tau$ as well.

To utilize the well-known M\"obius inversion, let us consider $\tau$ as an independent variable rather than a fixed scalar, Eq. (\ref{e02}) can be revised as
\begin{equation}\label{e05}
g(\tau,t)=f(\tau,t)+\tau\frac{D}{D t}f(\tau,t).
\end{equation}
Then a modified M\"obius series inversion theorem can be proposed as follows \cite{hardy,chen-1,nature,chen-2}: If
\begin{equation}\label{e06}
g(\tau,t)=\sum_{n=0}^{[\tau_{max}/{\tau}]}r(n)\tau^n\frac{D^n}{D t^n}f(\tau,t),
\end{equation}
then its exact inversion is given by
\begin{equation}\label{e07}
f(\tau,t)=\sum_{n=0}^{[\tau_{max}/{\tau}]}r^{-1}(n)\tau^n\frac{D^n}{D t^n}g(\tau,t),
\end{equation}
where the $r^{-1}(n)$ and $r(n)$ are satisfied

\begin{equation}\label{e08}
  \sum_{m+n=k}r^{-1}(n)r(m) =\sum_{n=0}^kr^{-1}(n)r(k-n)=\delta_{k,0}.
\end{equation}

Note that $\tau_{max}$ represents the maximum relaxation time under consideration, the change of the upper limit of the sums in Eq. (\ref{e06}) and Eq. (\ref{e07}) suggests that
\begin{equation}\label{e09}
f(\tau,t)|_{\tau>\tau_{max}}=0, \quad \mathrm{and }\quad g(\tau,t)|_{\tau>\tau_{max}}=0.
\end{equation}
Eqs. (\ref{e06}) and Eq. (\ref{e07}) are true under the condition of Eq. (\ref{e08}). Proof of Eq. (\ref{e07}) can be simply given below:

Substituting Eq. (\ref{e06}) into the right hand side of Eq. (\ref{e07}), we obtain
\begin{equation}\label{proof}
\begin{split}
&\sum_{n=0}^{[\tau_{max}/{\tau}]}r^{-1}(n)\tau^n\frac{D^n}{D t^n}g(\tau,t)\\
&= \sum \limits_{n=0}^{[\tau_{max}/{\tau}]}r^{-1}(n)\tau^n\frac{D^n}{Dt^n}\left(\sum_{m=0}^{[\tau_{max}/{\tau}]}r(m)\tau^m\frac{D^m}{D t^m}f(\tau,t)\right)\\
&= \sum \limits_{n=0}^{[\tau_{max}/{\tau}]}r^{-1}(n)\tau^n\left(\sum_{m=0}^{[\tau_{max}/{\tau}]}r(m)\tau^m\frac{D^{m+n}}{D t^{m+n}}f(\tau,t)\right)\\
&= \sum \limits_{n=0}^{[\tau_{max}/{\tau}]}\left(\sum_{m=0}^{[\tau_{max}/{\tau}]}r^{-1}(n)\tau^nr(m)\tau^m\frac{D^{m+n}}{D t^{m+n}}f(\tau,t)\right)\\
&= \sum \limits_{n=0}^{[\tau_{max}/{\tau}]}\left(\sum_{m=0}^{[\tau_{max}/{\tau}]}r^{-1}(n)r(m)\tau^{m+n}\frac{D^{m+n}}{D t^{m+n}}f(\tau,t)\right)\\
&= \sum \limits_{k=0}^{[\tau_{max}/{\tau}]}\left(\sum_{n=0}^{k}r^{-1}(n)r(k-n)\tau^{k}\frac{D^{k}}{D t^{k}}f(\tau,t)\right)\\
    &=\sum \limits_{k=0}^{[\tau_{max}/{\tau}]}\delta_{k,0}\tau^{k}\frac{D^{k}}{Dt^{k}}f(\tau,t)\\
     & =\tau^{0}\frac{D^{0}}{Dt^{0}}f(\tau,t)\\
    &=f(\tau,t).
\end{split}
\end{equation}

In order to apply the above result to Eq. (\ref{e05}), the function $r(n)$ must be
\begin{equation}\label{e10}
  r(n)=\left\{
              \begin{array}{cc}
                1 & n\leq1 \\
                0 & n> 1\\
              \end{array}
            \right.
\end{equation}
Based on Eq. (\ref{e08}), it can be obtained that
\begin{equation}\label{e11}
  r^{-1}(n)=(-1)^n.
\end{equation}
This means that the solution of Eq. (\ref{e05}) can be represented as
\begin{equation}\label{e12}
  f(\tau,t)=\sum \limits_{n=0}^{[\tau_{max}/{\tau}]}(-1)^n\tau^n\frac{D^n}{Dt^n}g(\tau,t).
\end{equation}
This is the modified Chapman-Enskog expansion which improves the traditional one given by Eq. (\ref{e04}).

It is easy to see that there is an essential difference between Eq. (\ref{e12}) and Eq. (\ref{e04}). The conventional summation in Eq. (\ref{e04}) is made up to infinity, however, the modified summation in Eq. (\ref{e12}) is made up to a finite number $[\tau_{max}/{\tau}]$. The infinite terms summation leads to divergent, however, the controllable finite terms of summation bring a convergent, since the number of summation is a function of the ratio of $[\tau_{max}/{\tau}]$. In the case of $\tau>\tau_{max}$, namely, $\tau_{max}/{\tau}<1$, both $g(\tau,t)$ and $f(\tau,t)$ will be no longer exist in Eq. (\ref{e12}), owing to the fact $[\tau_{max}/{\tau}]=0$ .

The modified Chapman-Enskog expansion in Eq. (\ref{e12}) can be viewed as an extension from a function \cite{hardy} to a differential operator.

Obviously, the new general solution in Eq. (\ref{e12}) can completely get rid of the divergence problem and is valid as long as $\tau$ is within $(0,\tau_{max}]$.

For example, set relaxation time range as $\tau \in [0.1,10]$, it is natural to take $\tau_{max}=10$. The solutions in Eq. (\ref{e12}) are changing as $\tau$ such as in Table \ref{tb1}
\begin{table}[h]
\caption{$f(\tau,t)=\sum \limits_{n=0}^{[\tau_{max}/{\tau}]}(-1)^n\tau^n\frac{D^n}{Dt^n}g(\tau,t)$}\label{tb1}
\footnotesize
\centerline{
\begin{tabular}{c|c|c|c}
\hline
$\tau$  &  $\frac{\tau_{max}}{{\tau}}$ & $[\frac{\tau_{max}}{{\tau}}]$  & Expression \\
\hline
$\tau=6$ & 1.67 & 1  & $f(\tau,t)=g(t)-\tau\frac{Dg}{Dt}$ \\
\hline
$\tau=4$ & 2.5 & 2 & $f(\tau,t)=g(t)-\tau\frac{Dg}{Dt}+\tau^2\frac{D^2g}{Dt^2}$\\
\hline
$\tau=2$ & 5 & 5  & $f(\tau,t)=g(t)-\tau\frac{Dg}{Dt}+\tau^2\frac{D^2g}{Dt^2}$\\
& & &$-\tau^3\frac{D^3g}{Dt^3}+\tau^4\frac{D^4g}{Dt^4}-\tau^5\frac{D^5g}{Dt^5}$\\
\hline
$\tau=1$ & 10 & 10 &  $f(\tau,t)=g(t)-\tau\frac{Dg}{Dt}+\tau^2\frac{D^2g}{Dt^2}$ \\
& & &$-\tau^{3}\frac{D^{3}g}{Dt^{3}}+\cdots +\tau^{10}\frac{D^{10}g}{Dt^{10}}$\\
\hline
$\tau=0.1$ &100 & 100  & $f(\tau,t)=g(t)-\tau\frac{Dg}{Dt}+\tau^2\frac{D^2g}{Dt^2}$ \\
& & &$-\tau^{3}\frac{D^{3}g}{Dt^{3}}+\cdots +\tau^{100}\frac{D^{100}g}{Dt^{100}}$\\
\hline
\end{tabular}}
\end{table}

The above solution is for a global range of relaxation time, if only "local" relaxation time is required, then $\tau_{max}=\tau$ could be taken, only two terms are involved in Eq. (\ref{e12}). The choice of $\tau_{max}$ should be determined by the physics of practical problems. However, the divergence problem has been overcome. The compatibility of Eq. (\ref{e12}) in all range of relaxation time indicates that there is a possibility to express a multi-scale flows in a unified way.

Having the modified Chapman-Enskog expansion in Eq. (\ref{e12}), we can have a new perspective on following scenarios:

\emph{Multi-scale scenario}: In gas dynamics \cite{landau}, in association with $\tau$, different scale of Knudsen number $Kn$ corresponding to different equations, for example, $Kn \sim 0$ relates to Euler's equation, $Kn < 0.01$ relates to Navier-Stokes equation without slipping boundary, $Kn \in [0.01,0.1]$ relates to Navier-Stokes equation with slipping boundary, $Kn \in [0.1,1]$  belongs to transition region and $Kn>10$ relates to Boltzmann equation for free-molecular motion. Back then, for different scale $\tau$ corresponding to different $Kn$, one has to adopt different Chapman-Enskog expansion. Now the solution in Eq. (\ref{e12}) provides a unified expression for multi-scale of relaxation time $\tau$.

\emph{Truncated series summation scenario}: In practical computations, it is very often to expand the first two terms in Eq. (\ref{e03}). It seems surprising that in this case the approximation solution is quite accurate with a reasonably large range of $\tau$.

For instance, if the $\tau_{max}$ is defined as $\tau \leq\tau_{max}< 2\tau$, or in ratio $1\leq\tau_{max}/\tau <2$, which leads to $[\tau_{max}/\tau]=1$, hence
\begin{equation}\label{e13}
\begin{array}{ccc}
  f(\tau,t)=\sum \limits_{n=0}^{[\tau_{max}/{\tau}]}(-1)^n\tau^n\frac{D^n}{Dt^n}g(\tau,t)\\
  =\sum \limits_{n=0}^{1}(-1)^n\tau^n\frac{D^n}{Dt^n}g(\tau,t)\\
  =g(\tau,t)-\tau \frac{Dg}{Dt}.
  \end{array}
\end{equation}

In conclusion, the highlights of this paper are listed in the Table \ref{tb2}.
\begin{table}[h]
\caption{Chapman-Enskog expansion and its modification}\label{tb2}
\footnotesize
\centerline{
\begin{tabular}{c|c}
\hline
Expansion  & Formulation  \\
\hline
Chapman-Enskog &  $f(t)=\sum \limits_{n=0}^\infty(-1)^n\tau^n\frac{D^n}{Dt^n}g(t)$ \\
\hline
Modified Chapman-Enskog & $f(\tau,t)=\sum \limits_{n=0}^{[\tau_{max}/{\tau}]}(-1)^n\tau^n\frac{D^n}{Dt^n}g(\tau,t)$  \\
\hline
\end{tabular}}
\end{table}

Eq. (\ref{e12}) could be considered as a more promising modification of the Chapman-Enskog expansion for future computations. It is an interesting coincidence that the present modification is proposed in the centenary of the Chapman-Enskog expansion. The application of this modified Chapman-Enskog expansion to other topics is remained for further study.

The authors wish to express their gratitude to Professor Chang Shu of National University of Singapore, who brings our attention to this problem. The constructive discussions with Professors Kun Xu of Hong Kong University of Science and Technology and Tao Tang of South China University of Science and Technology are also appreciated. The authors would like to express their most sincere thanks to reviewers for their high level academic inputs of comments and corrections. Their professionalism inspired us deeply. The permanent supports from State Key Laboratory of Low-dimensional Quantum Physics in Tsinghua University is also acknowledged .


\end{document}